\documentclass[
 aip,
 amsmath,
 amssymb,
 reprint,
 superscriptaddress,
 floatfix,
]{revtex4-1}

\usepackage{graphicx}
\usepackage{dcolumn}
\usepackage{bm}
\usepackage{mathrsfs}
\usepackage{xfrac}
\usepackage{hyperref}
\usepackage[mathlines]{lineno}
\usepackage{color}
\usepackage[normalem]{ulem}

\begin{document}

\title[Performance of a Large Area Photon Detector for Rare Event Search Applications]{Performance of a Large Area Photon Detector for Rare Event Search Applications}

\author{C.W.~Fink}\altaffiliation[]{Authors to whom correspondence should be addressed: \href{mailto:cwfink@berkeley.edu}{cwfink@berkeley.edu}, \href{mailto:samwatkins@berkeley.edu}{samwatkins@berkeley.edu}}\affiliation{Department of Physics, University of California, Berkeley, CA 94720, USA}
\author{S.L.~Watkins}\altaffiliation[]{Authors to whom correspondence should be addressed: \href{mailto:cwfink@berkeley.edu}{cwfink@berkeley.edu}, \href{mailto:samwatkins@berkeley.edu}{samwatkins@berkeley.edu}}\affiliation{Department of Physics, University of California, Berkeley, CA 94720, USA}
\author{T.~Aramaki}\affiliation{SLAC National Accelerator Laboratory/Kavli Institute for Particle Astrophysics and Cosmology, Menlo Park, CA 94025, USA}
\author{P.L.~Brink}\affiliation{SLAC National Accelerator Laboratory/Kavli Institute for Particle Astrophysics and Cosmology, Menlo Park, CA 94025, USA}
\author{J.~Camilleri} \altaffiliation[Now at ]{Department of Physics, Virginia Tech, Blacksburg, VA 24061, USA} \affiliation{Department of Physics, University of California, Berkeley, CA 94720, USA}
\author{X.~Defay} \affiliation{Physik-Department and Excellence Cluster Universe, Technische Universit\"at M\"unchen, 85747 Garching, Germany}
\author{S.~Ganjam} \altaffiliation[Now at ]{Department of Applied Physics and Yale Quantum Institute, Yale University, New Haven, CT 06511, USA}\affiliation{Department of Physics, University of California, Berkeley, CA 94720, USA}
\author{Yu.G.~Kolomensky} \affiliation{Department of Physics, University of California, Berkeley, CA 94720, USA}\affiliation{Lawrence Berkeley National Laboratory, Berkeley, CA 94720, USA}
\author{R.~Mahapatra}\affiliation{Department of Physics and Astronomy, and the Mitchell Institute for Fundamental Physics and Astronomy, Texas A\&M University, College Station, TX 77843, USA}
\author{N.~Mirabolfathi}\affiliation{Department of Physics and Astronomy, and the Mitchell Institute for Fundamental Physics and Astronomy, Texas A\&M University, College Station, TX 77843, USA}
\author{W.A.~Page} \affiliation{Department of Physics, University of California, Berkeley, CA 94720, USA}
\author{R.~Partridge} \affiliation{SLAC National Accelerator Laboratory/Kavli Institute for Particle Astrophysics and Cosmology, Menlo Park, CA 94025, USA}
\author{M.~Platt}\affiliation{Department of Physics and Astronomy, and the Mitchell Institute for Fundamental Physics and Astronomy, Texas A\&M University, College Station, TX 77843, USA}
\author{M.~Pyle}\affiliation{Department of Physics, University of California, Berkeley, CA 94720, USA}
\author{B.~Sadoulet}\affiliation{Department of Physics, University of California, Berkeley, CA 94720, USA} \affiliation{Lawrence Berkeley National Laboratory, Berkeley, CA 94720, USA}
\author{B.~Serfass}\affiliation{Department of Physics, University of California, Berkeley, CA 94720, USA}
\author{S.~Zuber} \affiliation{Department of Physics, University of California, Berkeley, CA 94720, USA}
\collaboration{CPD Collaboration}

\date{\today}

\begin{abstract}
We present the design and characterization of a large-area Cryogenic PhotoDetector (CPD) designed for active particle identification in rare event searches, such as neutrinoless double beta decay and dark matter experiments. The detector consists of a $45.6\,\mathrm{cm}^2$ surface area by 1-mm-thick $10.6\,\mathrm{g}$ Si wafer. It is instrumented with a distributed network of Quasiparticle-trap-assisted Electrothermal feedback Transition-edge sensors (QETs) with superconducting critical temperature $T_c=41.5\,\mathrm{mK}$ to measure athermal phonons released from interactions with photons. The detector is characterized and calibrated with a collimated $^{55}$Fe X-ray source incident on the center of the detector. The noise equivalent power is measured to be $1\times 10^{-17}\,\mathrm{W}/\sqrt{\mathrm{Hz}}$ in a bandwidth of $2.7\,\mathrm{kHz}$. The baseline energy resolution is measured to be ${\sigma_E = 3.86 \pm 0.04 \, (\mathrm{stat.})^{+0.19}_{-0.00} \, (\mathrm{syst.}) \, \mathrm{eV}}$~(RMS). The detector also has an expected timing resolution of ${\sigma_t = 2.3 \, \mu\mathrm{s}}$ for $5\,\sigma_E$ events.
\end{abstract}

\keywords{QET, Transition-Edge Sensor, dark matter, neutrino, phonon, photon}

\maketitle


In rare event searches, experimental sensitivity is often limited by background signals~\cite{PhysRevLett.120.132501, PhysRevLett.124.122501, Andreotti:2010vj, EdelweissWIMP, Angloher_2017, edelweissHV, Agnese_2018, Abramoff_2019, PhysRevLett.123.181802, Abdelhameed_2019}. Developing precision detectors to veto background and noise signals has been a high priority in these fields. Much interest in low temperature cryogenic detector technology has been shown by groups carrying out searches for neutrinoless double beta decay~\cite{RevModPhys.80.481} ($0\nu\beta\beta$), such as the CUORE~\cite{PhysRevLett.120.132501, PhysRevLett.124.122501}, CUPID~\cite{group2019cupid}, and AMoRE~\cite{Alenkov_2019} experiments. In these low-temperature calorimeters, the dominant source of background events consists of $\alpha$ decays from the surrounding environment~\cite{PhysRevLett.120.132501, PhysRevLett.124.122501, Azzolini:2019nmi}. It has been shown that Cherenkov emission or scintillation light can be used to positively identify the signal $\beta$s, allowing for background discrimination~\cite{beta_tag}. In order for these experiments to achieve a high level of rejection for these $\alpha$ backgrounds, photon detectors with large surface areas and baseline energy resolutions below $20\,\mathrm{eV}$ (RMS) for Cherenkov signals~\cite{beta_tag1}, or of $\mathcal{O}(100)\,\mathrm{eV}$ for scintillation signals~\cite{group2019cupid}, are required. To reject the pileup background from multiple ordinary (two neutrino) double beta decay ($2\nu\beta\beta$) events, experiments need timing resolutions down to $10\,\mu\mathrm{s}$ (for the $^{100}$Mo isotope)~\cite{group2019cupid}.

There has also been theoretical and experimental motivation to search for dark matter (DM) in the mass range of $\mathrm{keV}/c^2$ to $\mathrm{GeV}/c^2\,$~\cite{battaglieri2017cosmic, subgev,dark2013,dark2016}. However, current experiments have been limited by unknown background signals in the energy range of $\mathcal{O}$(1-100)$\,\mathrm{eV}$~\cite{EdelweissWIMP, Angloher_2017, edelweissHV, Agnese_2018, Abramoff_2019, PhysRevLett.123.181802, Abdelhameed_2019, PhysRevD.102.015017}. If the source of such backgrounds are high energy photons that deposit only an extremely small fraction of their energy in the target~\cite{PhysRevD.95.021301}, then a nearly $4\pi$ active shield composed of high-$Z$ scintillating crystals read out by these large area photon detectors could be highly efficient at suppressing these backgrounds. Additionally, a sensitive large area cryogenic detector could be useful for discriminating small energy depositions due to radiogenic surface backgrounds. Other potential DM applications for this detector technology include searches for inelastic electronic recoils off scintillating crystals and searches for interactions with superfluid He~\cite{PhysRevD.96.016026, KNAPEN2018386, PhysRevD.100.092007}.

We present the characterization of a large area Cryogenic PhotoDetector (CPD) with a measured baseline energy resolution of ${3.86  \pm0.04\,(\mathrm{stat.})^{+0.19}_{-0.00}\,(\mathrm{syst.})\,\mathrm{eV}}$~(RMS) and a timing resolution of $2.3\,\mu\mathrm{s}$ for $20\,\mathrm{eV}$ events that meets or exceeds the technical requirements for the currently proposed $0\nu\beta\beta$ experiments and DM searches. This is the first demonstration of the capabilities of such detectors, and further development may open opportunities for more novel applications.

\begin{figure}
    \includegraphics[width=\linewidth]{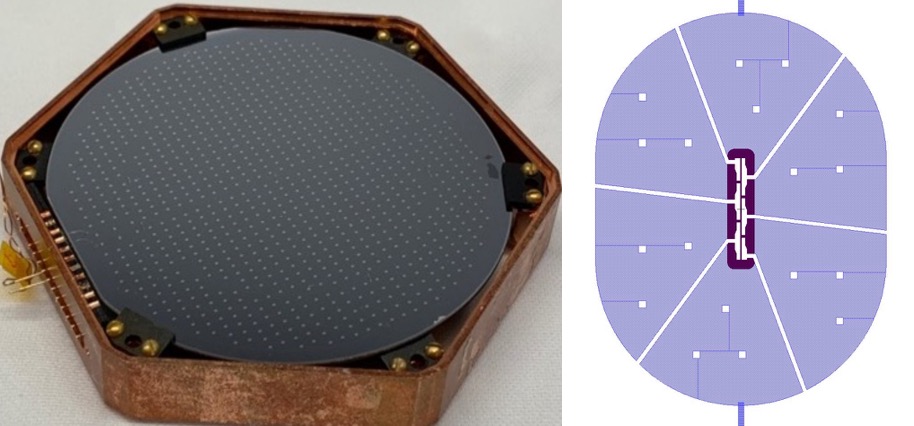}
    \caption{Left: A picture of the CPD installed in a copper housing. The instrumented side is shown facing up. Right: The design of the QETs used for the detector. (Blue: Al fins, Purple: W TES.)}
    \label{fig:det}
\end{figure}

The (100)-oriented substrate of the CPD is a $10.6\,\mathrm{g}$ Si wafer of thickness $1\,\mathrm{mm}$ and a surface area of $45.6\,\mathrm{cm}^2$. A parallel network of 1031 Quasiparticle-trap-assisted Electrothermal feedback Transition-edge sensors (QETs)~\cite{irwin,QET} with $T_c=41.5\,\mathrm{mK}$ was deposited on one side of the wafer. The QETs are uniformly distributed over the wafer's surface and connected to a single readout channel. The uniform and distributed nature of the channel allows for the fast collection of athermal phonons with minimal positional dependence, reducing efficiency penalties from effects such as athermal phonon down-conversion~\cite{knaak, downconversion}. The opposite side of the Si wafer is unpolished and noninstrumented. The detector and QET mask design can be seen in Fig.~\ref{fig:det}. In Table~\ref{tab:specs}, the QET design specifications for the CPD are listed.

\begin{table}
    \centering
    \caption{QET design specifications for the CPD describing the W TESs and the Al fins that each QET consists of. The active surface area refers to the amount of substrate that is covered by the Al fins of the QETs, while the passive surface area is that which is not covered by the Al fins, but by the Al bias rails, bonding pads, and other structures that absorb athermal phonons, but do not add to the signal.}
    \begin{tabular}{lc}
    \hline \hline
    Specification & Value \\ \hline
    TES Length $[\mu\mathrm{m}]$ & 140 \\
    TES Thickness $[\mathrm{nm}]$ & 40 \\
    TES Width $[\mu\mathrm{m}]$ & 3.5 \\
    Number of Al Fins & 6 \\
    Al Fin Length $[\mu\mathrm{m}]$ & 200 \\
    Al Fin Thickness $[\mathrm{nm}]$ & 600 \\
    Al-W Overlap $[\mu\mathrm{m}]$ & 10 \\
    Number of QETs & 1031 \\
    Active Surface Area $[\%]$ & 1.9 \\
    Passive Surface Area $[\%]$ & 0.2 \\ \hline \hline
    \end{tabular}
    \label{tab:specs}
\end{table}

The detector was studied at the SLAC National Accelerator Laboratory in a cryogen-free dilution refrigerator at a bath temperature ($T_B$) of $8\,\mathrm{mK}$. The detector was placed in a copper housing and was held mechanically with the use of six cirlex clamps. The cirlex clamps also provided the thermal link between the detector and the copper housing. The QET arrays were voltage biased and the current through the TES was measured with a DC superconducting quantum interference device (SQUID) array with a measured noise floor of ${\sim\!4~\mathrm{pA}/\sqrt{\mathrm{Hz}}}$.

A collimated $^{55}\mathrm{Fe}$ X-ray source was placed inside the cryostat and was incident upon the noninstrumented side of the CPD in the center of the detector. A layer of Al foil was placed inside the collimator to provide a calibration line from fluorescence at $1.5\,\mathrm{keV}$~\cite{alfluor, fe55}. The collimator was tuned such that there was $\sim\! 5\,\mathrm{Hz}$ of the $^{55}\mathrm{Fe}$ K$_\alpha$ and K$_\beta$ decays incident on the detector. The detector was held at a bath temperature ${T_B\ll T_c}$ for approximately two weeks to allow any parasitic heat added by the cirlex clamps to dissipate. During this time, we attempted to neutralize potential charged impurities within the Si wafer as much as possible with ionization produced by a $9.13\,\mu\mathrm{Ci}$ $^{137}\mathrm{Cs}$ source placed outside of the cryostat.


To characterize the QETs, $IV$ sweeps were taken at various bath temperatures by measuring TES quiescent current as a function of bias current\footnote{Although we are applying a bias current, we use the term ``$IV$'' because the voltage and current are related by the shunt resistor: ${V_{bias} = I_{\mathrm{bias}}R_{sh}}$}, with superimposed small square pulses providing complex admittance~\cite{irwin} at each point in the $IV$ curve~\cite{fink2020characterizing, Matt_thesis, Noah_thesis}. Since all the QETs are connected in parallel in a single channel, the channel was treated as if it were a single QET, describing the average characteristics of the total array. The $IV$ data allowed for the estimation of the parasitic resistance in the TES line ($R_p$), the normal state resistance ($R_N$), and the nominal bias power ($P_0$). The effective thermal conductance between the QETs to the Si wafer ($G_{TA}$) and $T_c$ were measured by fitting a power law to the measured bias power as a function of bath temperature~\cite{fink2020characterizing}. This measurement is a lower bound of these values, as it assumes no parasitic bias power in the system. We summarize these characteristics of the detector in Table~\ref{tab:rp}.

\begin{table}
    \centering
    \caption{Fitted and calculated parameters of the TES from $IV$ curves and complex impedance data. The complex impedance data are given for the bias point of $R_0\approx 35\%\, R_N$ (see Ref.~\onlinecite{fink2020characterizing} for definitions of parameters). The systematic errors on $G_{TA}$ and $T_c$ represent the upper bounds on these values, using the hypothesis that the observed excess noise in the sensor bandwidth is entirely due to parasitic bias power.}
    \begin{tabular}{ll}
    \hline \hline
    Parameter & Value \\ \hline
    $R_{sh}\, [\mathrm{m}\Omega]$ & $5\pm 0.5$ \\
    $R_p\, [\mathrm{m}\Omega]$ & $8.7\pm 0.8$ \\
    $R_N\, [\mathrm{m}\Omega]$ & $88 \pm 10$ \\
    $P_0\, [\mathrm{pW}]$ & $3.85 \pm 0.45$ \\
    $G_{TA}\, [\mathrm{nW}/\mathrm{K}]$ & $0.48 \pm 0.04\,(\mathrm{stat.})^{+0.49}_{-0.00}\,(\mathrm{syst.})$ \\
    $T_c\, [\mathrm{mK}]$ & $41.5\pm 1.0\,(\mathrm{stat.})^{+10}_{-0}\,(\mathrm{syst.})$ \\
    $R_0\,[\mathrm{m}\Omega]$ & $31\pm 3$ \\
    $\tau_0 \,[\mu\mathrm{s}]$ & $1700\pm 200 $ \\
    $L\, [n\mathrm{H}]$ & $190 \pm 10$\\
    $\beta$ & $1.1\pm 0.1$ \\
    $\mathscr{L} $ & $80\pm 15$ \\
    \hline \hline
    \end{tabular}
    \label{tab:rp}
\end{table}

\begin{figure}
    \includegraphics[width=\linewidth]{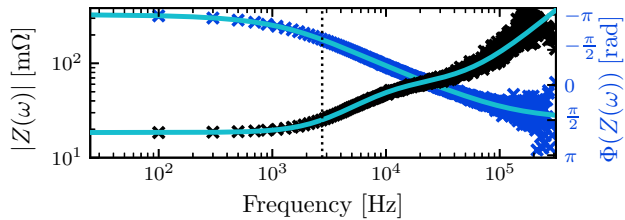}
    \caption{The magnitude and phase of the measured complex impedance are shown as the black and blue markers, respectively. The modeled complex impedance is shown as the cyan solid line. The black dotted line denotes the corresponding bandwidth of ${2.7\,\mathrm{kHz}}$ for the thermal time constant ${\tau_-=58\,\mu\mathrm{s}}$.}
    \label{fig:dvdi}
\end{figure}

The complex admittance data allows us to estimate the dynamic properties of the sensors. Throughout the superconducting transition, primary and secondary thermal fall times were observed, e.g. $58\,\mu\mathrm{s}$ and $370\,\mu\mathrm{s}$, respectively, at $R_0\approx 35\%\,R_N$. The origin of this additional time constant is under investigation. Its appearance suggests that we have a more complex thermal or electrical system, e.g. phase separation~\cite{2008JLTP..151...82C, Matt_thesis} or an extra heat capacity connected to the TES heat capacity~\cite{maasilta}. A characteristic plot of complex impedance of the TES circuit can be seen in Fig.~\ref{fig:dvdi}.

\begin{figure}
    \includegraphics[width=\linewidth]{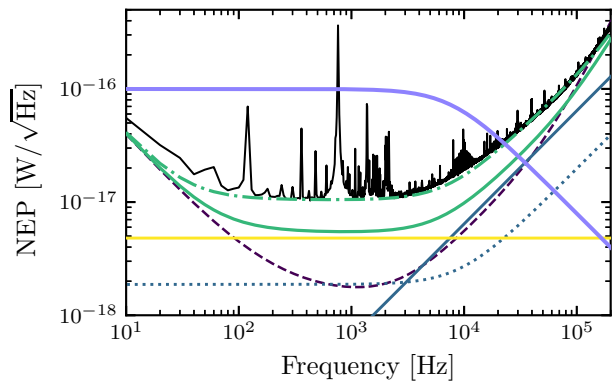}
    \caption{Modeled noise components: TES Johnson noise (blue solid), load resistor Johnson noise (blue dots), electronics noise (purple dashed), thermal fluctuation noise (TFN) between the TES and the bath (yellow solid), and total modeled noise (green solid) compared with the measured NEP (black solid) for $R_0 \approx 35\% R_N$. We additionally show the total noise model (green alternating dashes and dots), which includes a hypothetical environmental noise source of ${8\times 10^{-18}\, \mathrm{W}/\sqrt{\mathrm{Hz}}}$ and excess TES Johnson noise with $M=1.8$. The light-purple line in upper portion of the figure denotes the power-pulse shape (arbitrarily scaled), which consists of a single pole at the observed rise time of $1 / \left(2 \pi \tau_\mathrm{ph}\right)=8\,\mathrm{kHz}$.}
    \label{fig:transition_noise}
\end{figure}

Knowledge of the TES parameters, given in Table~\ref{tab:rp}, allowed for the calculation of the power-to-current responsivity, which was used to convert the measured current-referred power spectral density (PSD) to the noise equivalent power (NEP). These parameters were used to predict the expected noise spectrum using the single-heat-capacity thermal model~\cite{irwin}. A comparison of the NEP to the model at $R_0\approx 35\%R_N$ can be seen in Fig~\ref{fig:transition_noise}. The excess noise spikes above approximately $500\, \mathrm{Hz}$ have been experimentally confirmed to be largely caused by vibrations from the operation of the pulse tube cryocooler. The observed noise is also elevated above our model at frequencies in the effective sensor bandwidth interval (approximately the inverse of the thermal time constant~$\tau_-$~\cite{irwin}) by a factor of $\sim\!2$, as compared to the prediction. This ``in-band'' excess noise is consistent with two different hypotheses: a white power noise spectrum incident on the detector of $8\times 10^{-18}\, \mathrm{W}/\sqrt{\mathrm{Hz}}$ (e.g. a light leak) or a parasitic DC power in the bias circuit of approximately $6\,\mathrm{pW}$. If we assume the latter is the source, this allows us to calculate the upper bounds on our estimates of $G_{TA}$ and $T_c$, as reported in Table~\ref{tab:rp}. There remains bias-dependent excess noise above the sensor bandwidth. We parameterize the excess TES Johnson--like noise with the commonly used $M$ factor~\cite{irwin, doi:10.1063/1.3292343}. Using values of $M$ up to 1.8, depending on bias point, can account for the discrepancy between observation and prediction at these frequencies. We note that this ``excess'' noise could possibly also be explained with a more complex thermal model.

The lowest integrated NEP was achieved at an optimum bias point of ${R_0=31\,\mathrm{m}\Omega\approx 35\%R_N}$. In addition to the characterization data, approximately 500,000 threshold triggered events and 80,000 randomly triggered events were recorded at this bias.

For the measured phonon-pulse shape, there are multiple characteristic time constants. The pulse rise time was measured as ${\tau_{ph}=20\,\mu\mathrm{s}}$, which is the expected characteristic time scale for athermal phonons being absorbed by the Al collection fins of the QETs for this design. The dominant pulse fall time is consistent with the expectation from the complex impedance as we approach zero-energy, where we confirmed the expected thermal time constant ${\tau_-=58\,\mu\mathrm{s}}$ via a fit of the rise and fall times of the pulses. The secondary time constant from the complex impedance of $370\,\mu\mathrm{s}$ was also seen in these low-energy pulses. The secondary time constant from the complex impedance of $370\,\mu\mathrm{s}$ was also seen in these low-energy pulses, with an amplitude ratio of less than $2\%$ to the dominant decay exponential.

We observed an additional long-lived behavior in the pulses, which can be estimated as a low-amplitude $\sim\!3\,\mathrm{ms}$ exponential tail whose magnitude scales linearly with the event energy. As this tail is not seen in the complex impedance data, it might be due to direct absorption of phonons with energy smaller than the Al superconducting band gap into the TES~\cite{QET}.

\begin{figure}
    \includegraphics[width=\linewidth]{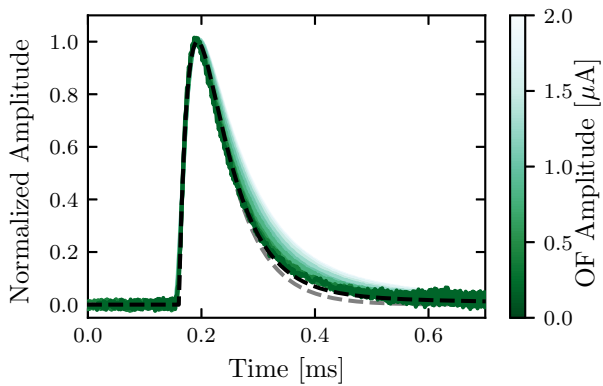}
    \caption{We show averaged pulse shapes (green solid) normalized by the peak current, for which the shade of green lightens with increased OF amplitude. For reference, $0.1\,\mu\mathrm{A}$ corresponds to about $0.1\,\mathrm{keV}$, $1.1\,\mu\mathrm{A}$ corresponds to about $1.5\,\mathrm{keV}$, and $2.0\,\mu\mathrm{A}$ corresponds to about $3.4\,\mathrm{keV}$. Each averaged pulse consists of about $100$ events averaged in $0.04\,\mu\mathrm{A}$ bin-widths. The lengthened fall time of the averaged pulse with increased OF amplitude (an energy estimator) is evident. The phonon-pulse template used in this analysis (black dashed) shows good agreement with the low energy (dark green) pulses. We also show an analytic phonon-pulse with only the first sensor fall time (gray dashed). Comparing to the phonon-pulse template, we see that the second sensor fall time has a small effect in this limited time interval.}
    \label{fig:pulse}
\end{figure}

For energies above $300\,\mathrm{eV}$, we observed a local saturation effect that manifests as the dominant fall time lengthening with increased energy. In Fig.~\ref{fig:pulse}, we show averaged pulses for various event amplitudes, showing the dependence of the pulse fall time on energy. We associate this effect with high-energy, single-particle events pushing nearby QETs into the normal resistance regime, slowing down the response of the total single-channel device. We also note that there is a position-dependent effect for a subset of high-energy events, notable by a varying fall time for events with the same amplitude. Our hypothesis for this phenomenon is that events close to the edge of the detector have less solid angle to deposit the energy, which leads to longer recovery times as opposed to events in the center of the detector (e.g. the calibration events). These effects are specific to the single-particle nature of the measured events. For scintillation events, the isotropic nature of the photons would spread out the event energy across the entire detector channel, avoiding these local saturation and position-dependent effects.

To reconstruct event energies, two energy estimators were used in this analysis: the optimum filter (OF) amplitude~\cite{OF,golwala} and the energy removed by electrothermal feedback ($E_{\mathrm{ETF}}$)~\cite{irwin}. For the OF, we used an offline algorithm to reconstruct energies. A single noise spectrum was used, which was computed from the randomly triggered events. The phonon-pulse template used was an analytic template that matches the measured low-energy pulse shape, neglecting the $3\,\mathrm{ms}$ low-amplitude tail. Because we could not directly measure the low-energy phonon-pulse shape with high statistics, we used a template without the long-lived behavior.

The integral estimator $E_{\mathrm{ETF}}$ was calculated for each triggered event by measuring the decrease in Joule heating via
\begin{equation}
    E_{\mathrm{ETF}} = \int_0^T \left[(V_b-2I_0R_\ell)\Delta I(t) - \Delta I(t)^2R_\ell\right]\mathrm{d}t,
    \label{eq:eabs}
\end{equation}
where $T$ is the time at which the integral is truncated, $\Delta I(t)$ is the baseline-subtracted pulse in current, $I_0$ is the quiescent current through the TES, $R_\ell$ is the load resistance, and $V_b$ is the voltage bias of the TES circuit~\cite{irwin}. In comparison to the OF amplitude, this integral estimator was less sensitive to saturation effects, but had a worse baseline energy resolution. When characterizing this device, we used the integral truncation of $T\approx7\tau_-$ for $E_{\mathrm{ETF}}$. This was done to preserve good baseline energy sensitivity in this integral estimator when calibrating the OF amplitude energy estimator at low energies.

\begin{figure}
    \includegraphics[width=\linewidth]{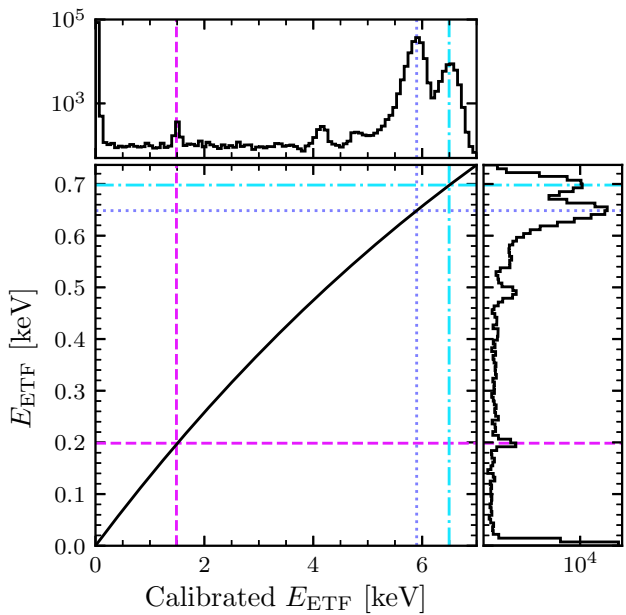}
    \caption{Upper: The calibrated $E_\mathrm{ETF}$ (which estimates $E_\mathrm{true}$) spectrum for the CPD (solid black) grouped in bins of width $70\,\mathrm{eV}$. Right: The energy spectrum in $E_\mathrm{ETF}$ (solid black) grouped in bins of width $7.4\,\mathrm{eV}$. Lower left: The fitted saturation model using Eq.~(\ref{eq:sat}) (solid black). In each of these panels, we have shown, for both the calibrated and uncalibrated $E_\mathrm{ETF}$ energy scales, the location of the K$_\alpha$, K$_\beta$, and Al fluorescence calibration peaks (pink dashed, blue dotted, and cyan alternating dashes and dots, respectively). In the lower left panel, the intersections of the lines corresponding to each spectral peak represent the points used for calibration of $E_\mathrm{ETF}$ via Eq.~(\ref{eq:sat}). The unmarked peaks at $4.2\,\mathrm{keV}$ and $4.8\,\mathrm{keV}$ in calibrated $E_\mathrm{ETF}$ are the Si escape peaks~\cite{Reed_1972}.}
    \label{fig:spectrum}
\end{figure}

For pulse-shape saturation at high energies, we use the following empirical model:
\begin{equation}
    E_{\mathrm{ETF}} = a \left(1 - \mathrm{exp}\left( - \frac{E_\mathrm{true}}{b}\right)\right).
    \label{eq:sat}
\end{equation}
This functional form has the expected behavior: it intercepts zero, approaches an asymptotic value at high energies, and becomes linear for small values of $E_{\mathrm{true}}$. In Fig.~\ref{fig:spectrum}, the fitted saturation model, as well as the calibrated and uncalibrated $E_{\mathrm{ETF}}$ spectra, are shown, as compared to the energies of various spectral peaks in both energy scales. For the event spectra, we observed an unknown background at low energies. As other surface experiments have seen excess backgrounds at similar energies~\cite{EdelweissWIMP, Angloher_2017}, we do not expect this to be detector-dependent. We are actively studying this detector at an underground facility, for which the results will be published in a future work.

The absolute phonon collection efficiency ($\varepsilon_{ph}$) of the detector was estimated by measuring $E_{\mathrm{ETF}}$ at the lowest energy calibration line (Al fluorescence) and dividing by the known energy of that line. Because of the long-lived behavior in the phonon-pulse shapes, the measured collection efficiency of this detector depends on the integration truncation time $T$. If it is chosen to only include energy collected by the first sensor fall time $\tau_-$ (e.g. ${T\approx7\tau_-}$), then we find that ${\varepsilon_{ph}=13\pm1\%}$. Alternatively, if we integrate to effectively infinity, this includes the low-amplitude long-lived behavior of the phonon pulses. In this case, the collection efficiency increases to ${\varepsilon_{ph}^\infty=17\pm1\%}$, which implies that about ${30\%}$ of the collected energy for a given event is associated with the low-amplitude tail of the phonon-pulse shape (about 8\% and 22\% from the $370\,\mu\mathrm{s}$ and $3\,\mathrm{ms}$ components, respectively).

To calibrate the OF amplitude to units of energy, we fit the relationship between the calibrated $E_{\mathrm{ETF}}$ and the OF amplitude to a linear slope at low energies (below approximately $300\,\mathrm{eV}$). This method does not provide a calibration of the OF amplitude at high energies, but allows for the calculation of the baseline energy resolution.

For the calibration method used, the main source of systematic error is the saturation model in Eq.~(\ref{eq:sat}). Since it is empirical, its use introduces uncertainty in its applicability. We can estimate the upper bound of the effect of this systematic on the baseline energy resolution as the value that would be reached if we instead calibrated $E_\mathrm{ETF}$ linearly using the Al fluorescence line. In this case, this worsens the baseline energy resolution, as we are not taking into account the expected response (see Fig.~\ref{fig:spectrum}).

The baseline energy resolution was calculated as the RMS of 46,000 randomly triggered events, after removing data contaminated by pileup events, electronic glitches, or thermal tails. This gave a resolution of ${\sigma_E=3.86\pm0.04\,(\mathrm{stat.})^{+0.19}_{-0.00}\,(\mathrm{syst.})\,\mathrm{eV}}$ (RMS) for the OF energy estimator, where these data are consistent with a normal distribution. This is in agreement with our estimation from the observed NEP and the power-referred phonon-pulse shape (a single-exponential with fall time $\tau_{ph}$ and collection efficiency $\varepsilon_{ph}$), which gave an expected baseline energy resolution of ${\sigma_E^{th}=3.9\pm 0.4\,\mathrm{eV}}$ (RMS), as was similarly done in Ref.~\onlinecite{fink2020characterizing}.

Using the OF formalism, we can also calculate the expected timing resolution~\cite{golwala} of the CPD, which provides an estimate of the minimum resolving time for two pileup events. For a $5\sigma$ event, the corresponding timing resolution of this detector is $2.3\, \mu\mathrm{s}$. For many $0\nu\beta\beta$ experiments, the minimum resolving time requirement to make pileup of multiple $2\nu\beta\beta$ events a negligible background is on the order of $1\,\mathrm{ms}$~\cite{Chernyak, artusa, pyle_CRESST, Casali_2019}. For the CUPID and CUPID-1T experiments, this requirement is about $300\,\mu\mathrm{s}$ and $10\,\mu\mathrm{s}$, respectively~\cite{group2019cupid}. An initial study of pileup events was carried out by adding two simulated $100 \, \mathrm{eV}$ pulses of randomized time separation to the in-run randomly triggered events from the CPD dataset. In this simulation, we observed that minimum resolving times below $10\, \mu\mathrm{s}$ are achievable with an OF-based pileup detection algorithm. In the future, we will study the minimum resolving time with a more detailed simulation based on the expected $2\nu\beta\beta$ spectrum for $^{100}$Mo. Given these initial studies, we expect the CPD to fulfill these requirements.

When comparing the baseline energy resolution of the CPD to the requirements of the CUPID experiment, the value surpasses the requirement of less than ${20\,\mathrm{eV}}$ (RMS) by a factor of five. While the CPD is a TES-based detector, it has been shown that Microwave Kinetic Inductance Detectors (MKIDs) and Neutron-Transmutation-Doped (NTD) Ge detectors are also promising avenues for achieving the sub-${20\,\mathrm{eV}}$ baseline goal. In Table~\ref{tab:comp}, we report this result alongside those of other detectors for this application. In comparison to the devices that have met or exceeded the requirement, the CPD does not require Neganov-Trofimov-Luke (NTL) amplification~\cite{Neganov:1985khw, doi:10.1063/1.341976} (which often results in excess dark counts) and has the best baseline energy sensitivity for its size.

\begin{table}
    \caption{Comparison of this work to various state-of-the-art devices for degraded $\alpha$ rejection in $0\nu\beta\beta$ experiments. The table is sorted by decreasing $\frac{\sigma_E}{\sqrt{\mathrm{Area}}}$, a common figure-of-merit of devices for this application. The column labeled ``NTL?" denotes whether or not each detector relies on NTL amplification to achieve the corresponding result.}
    \begin{tabular}{lcccc}
    \hline \hline
    \rule{0pt}{10pt} Device & Area $\left[\mathrm{cm}^2\right]$ & $\sigma_E$ $\left[\mathrm{eV}\right]$ & $\frac{\sigma_E}{\sqrt{\mathrm{Area}}}$ $\left[\frac{\mathrm{eV}}{\mathrm{cm}}\right]$ & NTL? \\ \hline
    MKID~\cite{Cardani_2018} & 4.0 & 26 & 13 & No \\
    W-TES~\cite{SCHAFFNER201530} & 12.6 & 23 & 6.5 & No \\
    Ge-NTD~\cite{BARUCCI2019150} & 15.6 & 20 & 5.1 & No \\
    Ge-NTD~\cite{Pattavina_2015} & 19.6 & 19 & 4.3 & Yes \\
    IrAu-TES~\cite{Willers_2015} & 4.0 & 7.8 & 3.9 & Yes \\
    Ge-NTD~\cite{Armengaud_2017}& 4.9 & 7.6 & 3.5 & Yes \\
    Ge-NTD~\cite{PhysRevC.97.032501} & 15.2 & 10 & 2.6 & Yes \\
    Ge-NTD~\cite{Novati_2019} & 15.2 & 8 & 2.1 & Yes \\
    W-TES~\cite{2018JLTP..193.1160R} & 12.6 & 4.1 & 1.2 & No \\
    W-TES (this) & 45.6 & 3.9 & 0.6 & No \\ \hline \hline
    \end{tabular}
    \label{tab:comp}
\end{table}


The measured baseline energy resolution of ${3.86\pm0.04\,(\mathrm{stat.})^{+0.19}_{-0.00}\,(\mathrm{syst.})\,\mathrm{eV}}$ and the expected timing resolution of ${2.3\, \mu\mathrm{s}}$ (at $5\sigma_E$), combined with its large surface area, makes this detector an excellent candidate for background rejection in both $0\nu\beta\beta$ and DM experiments. Because of the energy sensitivity, this device can be used as a dark matter detector itself, as we have done in collaboration with SuperCDMS to set limits on spin-independent dark matter-nucleon interactions for sub-$\mathrm{GeV}/c^2$ dark matter particle masses~\cite{alkhatib2020light}. Similarly, this gram-scale device could be applied to coherent elastic neutrino-nucleus scattering experiments~\cite{coh}. The performance of the CPD can be further optimized through adjustment of characteristics such as the Al-W overlap and overall Al coverage. From these considerations, we anticipate up to a factor of two improvement in baseline energy resolution for a future iteration of the CPD, which is currently being designed.


Authors C.W.F. and S.L.W. contributed equally to this work. This material is based upon work supported by the US Department of Energy (DOE) Office of Science under Contract Nos. DE-AC02-05CH11231 and DE-AC02-76SF00515, by the DOE Office of Science, Office of High Energy Physics under Contract Nos. KA-2401032, DE-SC0018981, and DE-SC0017859, by the National Science Foundation (NSF) under Grant Nos.  PHY-1314881, PHY-1415388, and PHY-1809769, and by Michael M. Garland.


The data that support the findings of this study are available upon reasonable request to the corresponding authors.


\bibliographystyle{aipnum4-1}
\bibliography{cpd_technical_arxiv}

\end{document}